\documentclass[vecphys]{svmult}

\usepackage{makeidx}         
\usepackage{graphicx}        
\usepackage{multicol}        
\usepackage[bottom]{footmisc}

\makeindex             

\def\nuebar{\bar{\nu_e}}
\def\dm2{\rm{\Delta m^2}}

\begin{document}


\title*{Neutrino Physics and Astrophysics : Highlights}
\author{Henry T. Wong\inst{1}}
\institute{Institute of Physics, Academia Sinica, Taipei 11529, Taiwan.
\texttt{htwong@phys.sinica.edu.tw}}
%
%

\maketitle

\hfill AS-TEXONO/07-02 \\
\hspace*{1cm} \hfill \today

\vspace*{0.5cm}


\begin{flushleft}
{\bf Abstracts}\\
\end{flushleft}

This article
presents an overview of neutrino
physics research, with
highlights on the physics goals,
results and interpretations
of the current neutrino
experiments and future directions
and program.

\section{Introduction}
\label{sec:1}

One of the major breakthrough in
elementary particle physics
in the past decade is the evidence of neutrino
masses and mixings through the studies of
neutrino oscillations.
It offers the first case for the need to extend
the much-cherished Standard Model.
Several important questions are raised, and
there  are intense
world-wide efforts to pursue the next-generation
of neutrino experiments to address them.

The objective of this
article is to ``set the stage''
for students and researchers not in the field
by summarizing the key ingredients and highlights
of the goals, status and future directions in
neutrino physics.
It is not meant to be a comprehensive lecture or
detailed review article.
Interested readers can pursue the details via
the listed references in
the Review of Particle Physics~\cite{pdg}
textbooks~\cite{book},
conference proceedings~\cite{nu04}
and Web-links~\cite{weblink}.

\section{Neutrino Physics : WHY }

Neutrino exists $-$ and exists in large quantities in
the Universe, comparable in number density to the photon.
It is known that there are three flavor
of light neutrino coupled
via weak interaction to the Z gauge boson.
Similar to the quark sector, the flavor (or interaction)
eigenstates $\nu _l$ is a linear combination of 
the mass eigenstates $\nu _i$ via a mixing matrix {\it U},
usually called the PMNS matrix 
(after Pontecorvo-Maki-Nakagawa-Sakata).
Denoting the flavor and mass eigenstates respectively by
\begin{equation}
\label{eq::def}
\Psi ~ \equiv ~ 
\left( \begin{array}{c} \nu _e \\
                        \nu _{\mu} \\
                        \nu _{\tau}  \end{array} \right)
~ ~~ ~ \&  ~ ~~ ~
\phi ~ \equiv ~ 
\left( \begin{array}{c} \nu _1 \\
                        \nu _2 \\
                        \nu _3  \end{array} \right)  ~~ ,
\end{equation}
one have
\begin{equation}
\label{eq::mixing}
\Psi ~ = ~
U ~ \phi ~~ .
\end{equation}
Unitarity constraints 
imply that {\it U} can be described 
four independent variables which 
are usually parametrized as:
\begin{equation}
\label{eqn::u}
U ~ \equiv ~ U ~
( ~ \theta_{12} ~,~ \theta_{23} ~,~ \theta_{13} ~,~ \delta ~ )
\end{equation}
where $\theta_{ij}$ are
the mixing angles 
between $\nu_i$ and $\nu_j$,
and $\delta$ is a phase that characterizes possible
CP-violation.
Being electrically-neutral, the neutrinos can be
``Dirac'' or ``Majorana'' particles, identified
by whether the neutrinos and anti-neutrinos
are different or the same, respectively.
For Majorana neutrinos, a diagonal matrix 
\begin{equation}
U_{Maj} ~ = ~  diag ~ ( ~ 1 ~  , ~ e^{i \alpha} ~ , ~ e^{i \beta} ~ )
\end{equation}
should be added such that the mixing matrix becomes
$U \rightarrow U U_{Maj}$.

\begin{figure}
\centering
\includegraphics[width=9cm,angle=270]{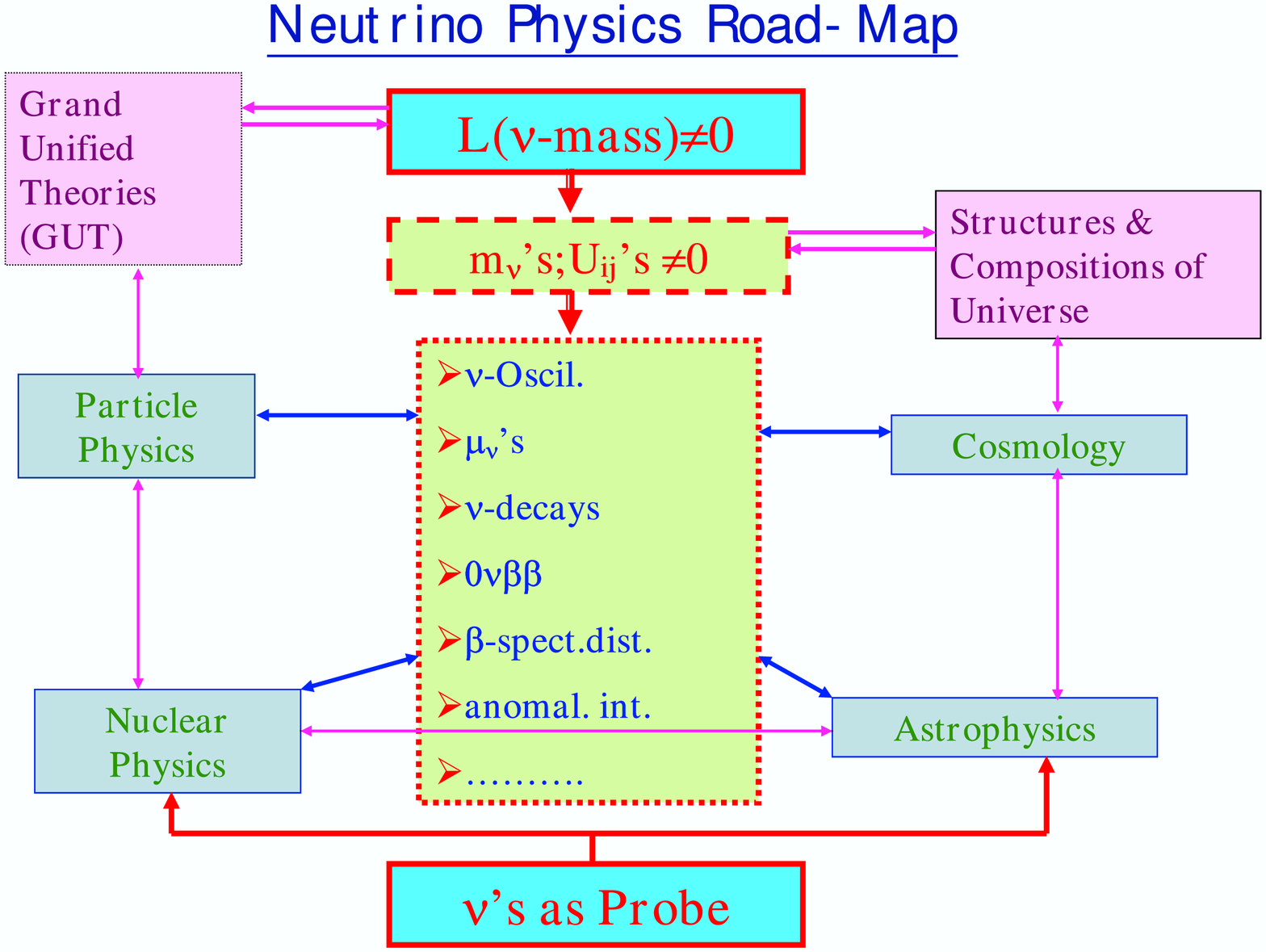}
\caption{
A schematic chart showing the
relationships between different
domains in neutrino physics research.
}
\label{roadmap}       
\end{figure}

The ``road map'' in neutrino physics research
can be summarized in Figure~\ref{roadmap}.
In field theory language, 
the experimentally measurable neutrino properties 
can be encoded 
in a ``neutrino mass term'' $L ( m_{\nu} )$
in the Lagrangian. Standard Model sets this to be identically
zero, but without any compelling reasons $-$ in
contrast to the massless-ness of the photons being dictated
by gauge invariance.
The detailed structures and values
of this term can reveal much about
the Grand Unified Theories.

The mass term can be expressed either as:
\begin{equation}
\label{eqn::psi}
 L ( m_{\nu} )  ~ = ~ \bar{\Psi}_R  \cdot M_{3X3} \cdot \Psi_L
\end{equation}
in the flavor basis or as
\begin{equation}
\label{eqn::phi}
 L ( m_{\nu} ) ~ = ~ \bar{\phi}_R  \cdot M_{D} \cdot \phi_L
\end{equation}
in the mass basis, where
the 3$\times$3 matrix $M_{3X3}$ is related to
the diagonalized mass matrix $M_D$ by
\begin{equation}
M_{3X3} ~ = ~
U^T \cdot M_D \cdot U ~~ .
\end{equation}
The fundamental intrinsic neutrino
properties of great interest
are the ``neutrino masses'' 
$m_i$ of $\nu_i$ (for $i=1,2,3$) 
given in 
\begin{equation}
U_D ~ = ~  diag ~ ( ~ m_1 ~ , ~ m_2 ~ , ~ m_3 ~ )
\end{equation}
and the ``neutrino mixings'' due to
individual elements 
$U_{li}$ of {\it U}.
The $m_i$ and $U_{li}$
are experimentally measurable, from which
$M_{3X3}$ can be constructed.
The high energy new physics scale 
as well as parameters for 
symmetry breaking and mass generation
are expected to be leave their imprints in $M_{3X3}$.
Accordingly, the pursuit of neutrino masses and mixings
will lead to a knowledge of $M_{3X3}$ which will shed
light to the deep-hidden secrets of nature.

At the large length-scale frontier,  
neutrino mass is related to
the composition and structural evolution of
the Universe. 
The combined neutrino and cosmology 
data~\cite{pdgcosmos} indicate
that our Universe is at critical density 
and that neutrinos 
constitute to at least 0.1\% of this density,
comparable to the fraction shared by visible
matter.

\begin{figure}
\vspace*{0.7cm}
\centering
\includegraphics[width=10cm]{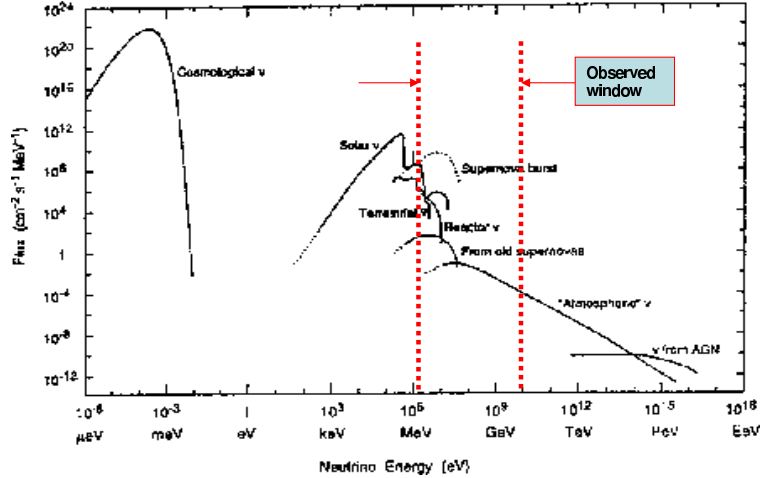}
\caption{
The expected neutrino spectra from
various celestial and terrestrial sources.
Neutrinos from man-made accelerators, typically
at the range of 10~MeV to 100~GeV, are not
shown here since accelerator parameters differ.
The detectable window is also shown.
}
\label{nusource}
\end{figure}

The study of 
neutrino physics and the implications
of the results connect many disciplines together,
from particle physics to nuclear physics to astrophysics
to  cosmology.
Experimentally, the probing of  $L ( m_{\nu} )$ 
is carried out by studying various processes
related to neutrino masses and mixings, such as direct mass
measurement through the distortion of $\beta$-spectra,
neutrinoless double beta decays, neutrino oscillations,
neutrino magnetic moments, neutrino decays ..... and
so on. These investigations are realized by a wide
spectrum of experimental techniques spanned over many
decades of energy scale with different neutrino sources. 
The expected neutrino spectrum due to terrestrial
and astrophysical sources are shown in Figure~\ref{nusource}.

In addition, neutrino has been used as probe (as ``beam''
from accelerators and reactors and even astrophysical
sources) to study electroweak physics,
QCD, structure function physics,  nuclear physics,
and to provide otherwise inaccessible information on
the interior of stars and supernovae.

Neutrino interactions are characterized by cross-sections
at the weak scale (100~fb at 100~GeV to $< 10^{-4}$~fb      
at 1~MeV). As an illustration, the mean free path in water
for $\nuebar$ from power reactors at the typical
energy of 2~MeV  is 250 light years!
The central challenge to neutrino experiments is
therefore on how to beat this small cross section.
The typical solution is to deploy
massive detectors. 
Consequently the experimental
hurdles become how to keep the cost and 
background manageable. These pose great challenges
and demand
to the creativity $-$ and often courage $-$
of experimentalists.

\section*{Neutrino Physics :  NOW }

After half a century of ingenious experiments
since the experimental discovery of the neutrinos
by Cowan and Reines~\cite{reinescowan},
there are several results which indicate compellingly
the existence of neutrino masses~\cite{kayserpdg,vogelpdg}, 
and hence
the necessity for the extension of
the Standard Model.  All these
results are based on experimental studies
of neutrino oscillation, a 
quantum-mechanical interference effect which 
allows neutrino to transform from one flavor
eigenstate to another as it traverses in space.

\begin{figure}
\centering
\includegraphics[width=12cm]{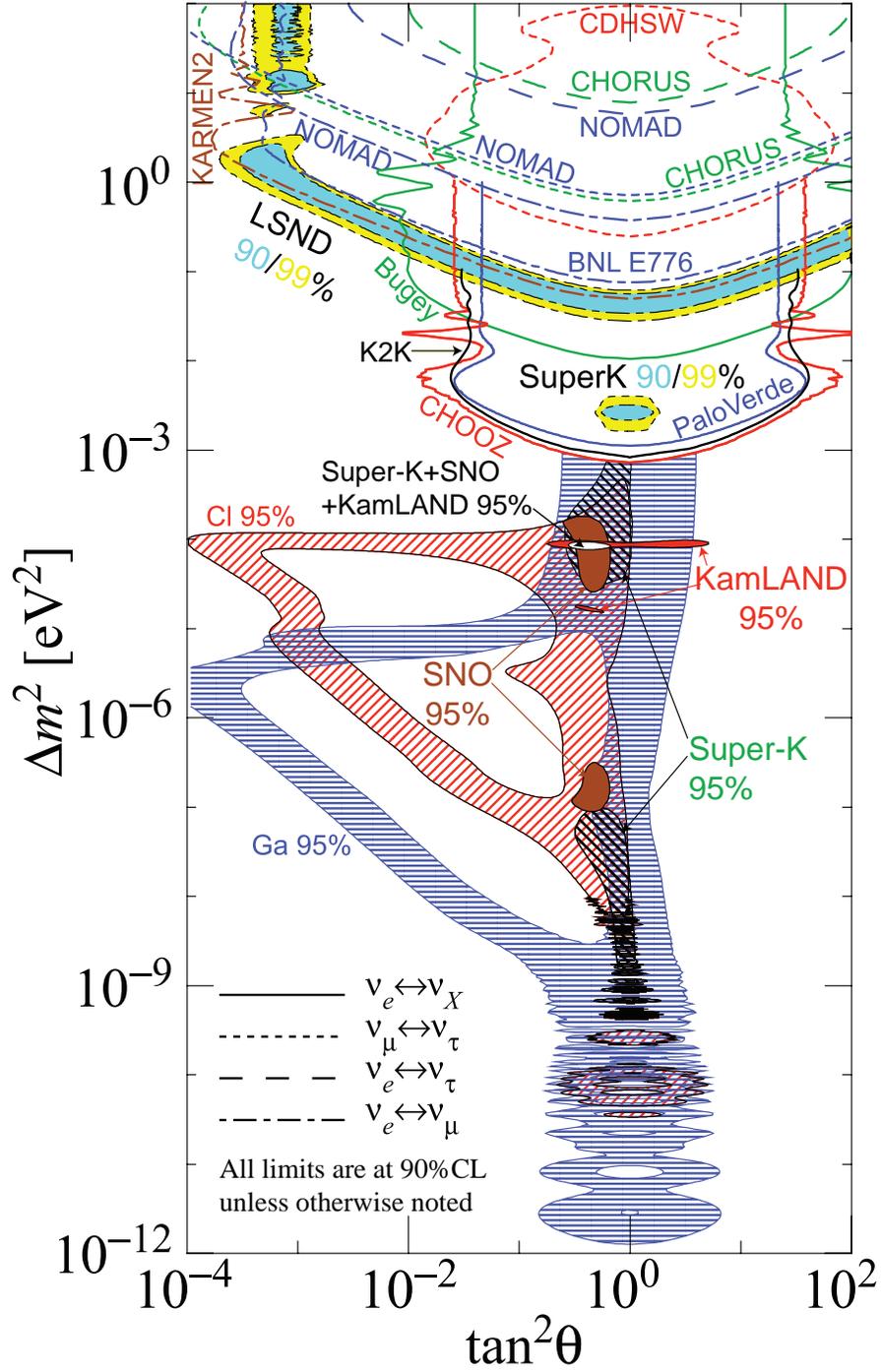}
\caption{ 
A summary of results of neutrino oscillation
experiments in ($\Delta m^2$, $\theta$)
parameter space.
The allowed regions from different
experiments are shown by various
shading. 
Adapted from Ref.~\cite{kayserpdg}.
}
\label{fig::nuosc}
\end{figure}

Using the two-flavor formulation of the
two lighter families as illustration,
Eq.~\ref{eq::mixing} can be written as 
\begin{equation}
\left( \begin{array}{c} \nu _e \\
                        \nu _{\mu} \end{array} \right)
~ = ~
\left( \begin{array}{cc} cos ~ \theta _{12} ~ & ~ sin  ~ \theta _{12} \\
                        - sin ~ \theta _{12} ~ & ~ cos ~ \theta _{12} 
			\end{array} \right)  ~
\left( \begin{array}{c} \nu _1 \\
                        \nu _2 \end{array} \right)  ~~ .
\end{equation}
The probability of $\nu _e$ at energy $E_{\nu}$
transforming to $\nu _{\mu}$ 
after traversing a distance $L$ is 
given by
\begin{equation}
P ( \nu _e \rightarrow \nu _{\mu} ) 
=
sin ^2 2 \theta  ~ sin ^2 ( \frac{\Delta m_{12} ^2 ~ L }{E_{\nu}} )
\end{equation}
where $\Delta m_{12} ^2 =  m_2^2 - m_1^2 $ 
is the mass-squared difference
between the two mass eigenstates.
The ($E_{\nu} , L$) parameters are select-able in 
individual experiments such that
different regions of ($\Delta m^2 , \theta$) are
probed.
In the presence of matter, a resonance ``MSW''
effect~\cite{msw} can take place when the condition
\begin{equation}
\sqrt{2} \cdot G_F \cdot \rho_e ~ = ~ 
\frac{\Delta m^2}{2 E_{\nu}} \cdot  
{\rm cos} 2 \theta
\end{equation}
are satisfied, where 
$G_F$ is the Fermi constant and 
$\rho_e$ is the electron density in matter. 
The effective mixings at resonance are maximal 
even though the vacuum mixing angle $\theta$ can be small.
The origins are due to the physics that
($\nu_{\mu} ,  \nu_{\tau}$) only interact via
neutral currents with the electrons while 
$\nu_e$ can have both neutral and charged-current
interactions.
A summary of the results
of neutrino oscillation experiments
is shown in Figure~\ref{fig::nuosc}.

Our current knowledge of the mixing matrix {\it U}
and of $\Delta m^2$ is derived from positive
evidence of neutrino masses and mixings from
several experiments using 
different neutrino sources:

\subsection{Atmospheric and Accelerator Neutrinos}

Data from the Super-Kamiokande experiments~\cite{superk},
supported by other experiments (MACRO, SOUDAN),
indicate a smaller  
$\rm{ ( \nu_{\mu} + \bar{\nu_{\mu}} ) /
( \nu_{e} + \bar{\nu_{e}} ) }$ 
ratio
than would be expected from propagation
models of cosmic-ray showering. 
There is 
a statistically strong
dependence of the deficit
with the zenith angle of the neutrinos,
which can be translated to
a dependence with $L / E_{\nu}$,
a ``smoking gun'' evidence for neutrino
oscillations.
The data supports a scenario
of $\rm{\nu_{\mu} \rightarrow \nu_{\tau}}$
oscillation, rather than $\nu_{\nu}$ disappearing
into sterile neutrinos.

The allowed region is indicated by ``SuperK''
in the ($\Delta m^2 , \theta$) plot of
Figure~\ref{fig::nuosc}, with
\begin{eqnarray}
\Delta m^2_{atm} ~ & = & 
( 1.9 - 3.0 ) \times 10^{-3} ~ {\rm eV ^2} \nonumber \\
sin ^ 2 2 \theta_{atm} ~ & > & 0.9  
\label{eqn::atm}
\end{eqnarray}

Long baseline neutrino oscillation experiment K2K~\cite{k2k} 
compared the flux of 
an accelerator $\nu_{\mu}$ beam 
close to its production point 
at the KEK Laboratory to that measured 
at the Super-Kamiokande detector
after the beam traversed a distance of 250~km. 
Disappearance of the $\nu_{\mu}$ was observed
at a level compatible with the 
($\Delta m^2_{23} , \theta_{23}$)
measurements of the atmospheric neutrino
experiments.

\subsection{Solar and Reactor Neutrinos}

All previous solar neutrino experiments 
(Homestake, Kamiokande, GALLEX, SAGE, Super-Kamiokande)
observed solar neutrino flux less
than the predictions of Standard Solar
Model~\cite{bahcallbook}. 
The deficit is different
among the experiments, suggesting an
energy dependence of the effect. 

\begin{figure}
\centering
\includegraphics[width=10cm]{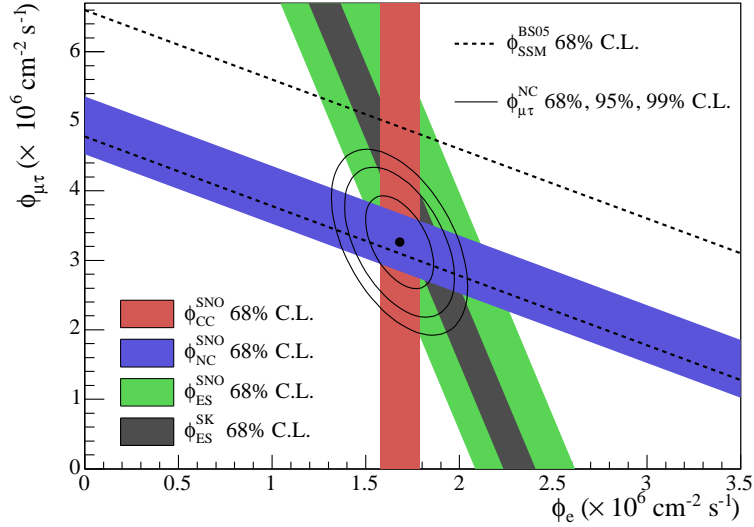}
\caption{
Results from the SNO experiment showing
the complementary regions of sensitivities
from the charged-currents (CC), neutral currents (NC)
and electron scattering (ES) measurements. 
Predictions by the Standard Solar Model are 
the region between the dotted lines.
Adapted from Ref.~\cite{sno}.
}
\label{snoplot}
\end{figure}

The SNO experiment~\cite{sno} in Canada,
with its deuteron target in the form
heavy water, measured the solar neutrino
flux in three different channels.
As depicted in Figure~\ref{snoplot},
the flavor-dependent
elastic scattering (ES) and charged-current (CC)
channels shows deficit while the 
flavor-independent
neutral-current channel agrees
perfectly with the predictions from the
standard Solar Model~\cite{bahcallbook,solarmodel}.
These represent compelling evidence that
neutrinos are produced in the correct amount
by nuclear fusion in the interior of the Sun,
and that some $\nu_{e}$'s produced 
are converted to another active neutrino flavors
on its way to Earth.  The 
``Solar Neutrino Problem'', which has been
a major puzzle in basic science for thirty
years, is thus solved.

The combined solar neutrino data 
can be explained by neutrino oscillation
in matter and select
the ``Large Mixing Angle'' (LMA) region
as labeled by ``SNO''
in Figure~\ref{fig::nuosc} as the
preferred solution. 

The KamLAND~\cite{kamland} experiment 
in Kamioka, Japan is sensitive to the 
$\nuebar$ emitted from the many
nuclear power reactors in the vicinity,
with an average distance of 180~km.
An energy-dependent deficit of the 
$\nuebar$ flux was observed.
When interpreted as neutrino oscillation
in vacuum, the allowed parameter space as indicated
by ``KamLAND'' in Figure~\ref{fig::nuosc}
has excellent overlap with the solar
neutrino LMA region and significantly
enhance the sensitivity to $\Delta m^2$.
The combined fit of the solar and
reactor neutrino data gives:
\begin{eqnarray}
\Delta m^2_{\odot} ~ & = & 
8.0 \pm ^{0.6}_{0.4} \times 10^{-5} ~ \rm{ eV^2} \nonumber \\
tan ^2 \theta_{\odot} ~ & = & 0.45 \pm ^{0.09}_{0.07}
\label{eqn::solar}
\end{eqnarray}

\subsection{LSND Anomaly:}
The LSND experiment~\cite{lsnd} with accelerator neutrinos
reported unexpected excess of $\rm{\nu_{e}}$ 
in a 
$\rm{  \nu_{\mu} + \bar{\nu_{\mu}} }$ beam,
which can be explained by 
$\rm{\nu_{\mu} \rightarrow \nu_{e}}$
oscillation at the relative large $\Delta m^2$
and small $\theta$ region as denoted by
``LSND'' in Figure~\ref{fig::nuosc}.
The results are yet to
be reproduced (or totally excluded)
by other experiments.
The MiniBoone experiment, currently
under data taking and analysis, will
confirm or refute these anomalous results.

\subsection{Neutrino Masses and Mixings}

If one takes the conservative approach
that the LSND results must be confirmed by an
independent experiment
before they are incorporated into
the theoretical framework, then 
a three-family scheme with neutrino
masses and mixings is adequate.
Otherwise, sterile neutrinos must be
added and the Standard Model must be
substantially extended.

\begin{figure}
\centering
\includegraphics[width=10cm]{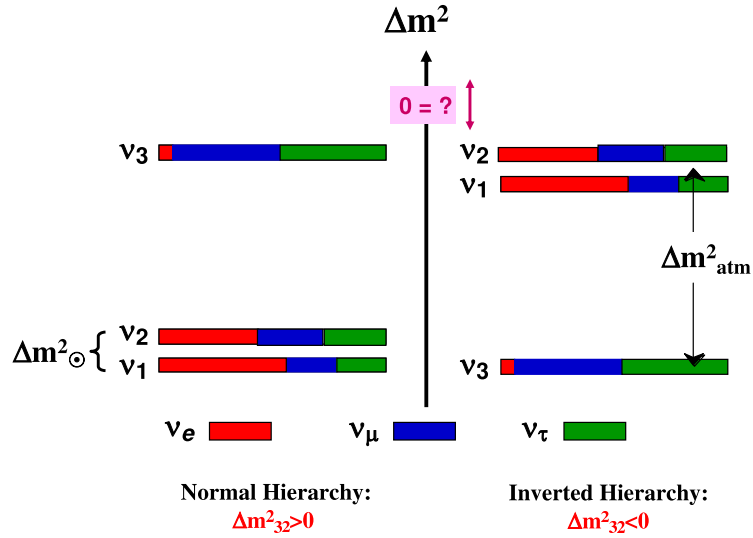}
\caption{
The favored neutrino mass spectra 
to explain the various positive 
results  from neutrino oscillation
experiments.
}
\label{mhier}
\end{figure}

Accordingly, 
the atmospheric and solar neutrino results
given in Eqs.~\ref{eqn::atm} and \ref{eqn::solar}
are due to $\nu_2 \leftrightarrow \nu_3$
and $\nu_1 \leftrightarrow \nu_2$
oscillations, respectively:
\begin{eqnarray}
\Delta m^2_{23} & = & \Delta m^2_{atm} ~ ; ~ 
\theta _{23}  =  \theta _{atm}  \\
\Delta m^2_{12} & = & \Delta m^2_{\odot} ~ ; ~ 
\theta _{12}  =  \theta _{\odot}  
\end{eqnarray}
The oscillation
and $\nu_1 \leftrightarrow \nu_3$
is not observed experimentally yet, and
only upper limit from reactor experiment
CHOOZ~\cite{chooz} exists for $\theta _{13}$:
\begin{eqnarray}
\Delta m^2_{13} ~ \sim ~ 
\Delta m^2_{23} ~ & = & 
( 1.9 - 3.0 ) \times 10^{-3} ~ {\rm eV ^2} \nonumber \\
sin ^2 \theta_{13} ~ & <  & 0.05
\label{eqn::13}
\end{eqnarray}

Taking $\theta _{12} = 45^o$,
the mixing matrix {\it U} in Eq.~\ref{eqn::u}
has the structure:
\begin{equation}
U \sim 
\left( \begin{array}{ccc} 
 cos \theta _{12} & sin \theta _{12} & e ^{i \delta} sin \theta _{13} \\
- sin \theta _{12} / \sqrt{2} & cos \theta _{12} / \sqrt{2} & 1 / \sqrt{2} \\
sin \theta _{12} / \sqrt{2} & - cos \theta _{12} / \sqrt{2} & 1 / \sqrt{2} 
\end{array} \right) ~ ,
\end{equation}
where $\theta _{12} \sim 34 ^o$.
The mixing angle 
$\theta _{13}$ and the CP-violating phase $\delta$
remain unknown.
The matrix $U_{Maj}$ is diagonal and hence is not
sensitive to oscillation experiments.
An important feature is that while the formalism
are almost identical, the structure of the
mixing matrix in
the neutrino sector is very different from
that in the quark sector.
In particular, cases where $\theta _{23} = 45^o$
or $\theta _{13} = 0$ exactly
may imply symmetry principles 
in nature yet to be discovered.

Our present knowledge of 
the neutrino mass spectrum is depicted 
schematically in Figure~\ref{mhier}.
The color-codes denote the
compositions of the various mass
eigenstates.
Depending on which is the 
eigenstate with the minimum mass,
there are two possibilities: the normal
and the invert hierarchies.
Since the oscillation results only provides
information on $\Delta m^2$'s, 
the absolute scale of $m_i$ remains
unknown. It can be ``hierarchical'',
in which 
$m_i (max) \sim \sqrt{\Delta m^2_{23}} \sim 45~{\rm meV}$,
or ``quasi-degenerate'', where
$m_1 \sim m_2 \sim m_3$.
Existing $\beta$-spectrum measurements
give upper bound of $m _i < 2 ~ \rm{eV}$ for
the individual neutrino masses.

\section{Neutrinos in Cosmology}

There is a long tradition of interplay
between neutrino physics and cosmology
$-$ that is, physics at the smallest
and largest scale.
The standard model of
primordial nucleosynthesis~\cite{pdgbbn}
is a pillar of the Big-Bang Cosmology.
Comparisons of the 
observed baryon-to-photon ratio ($\eta$)
in the Universe
with the abundance of $^4$He, D, $^3$He and $^7$Li 
can place constraints on
the degrees of freedom on their production,
and hence the number of light neutrino families.
The latest global analysis,
including a determination of $\eta$ from 
the cosmic microwave background measurements,
gives $N_{\nu} = 3.24 \pm 1.2$~\cite{nufamily}.

\begin{figure}
\centering
\includegraphics[width=10cm]{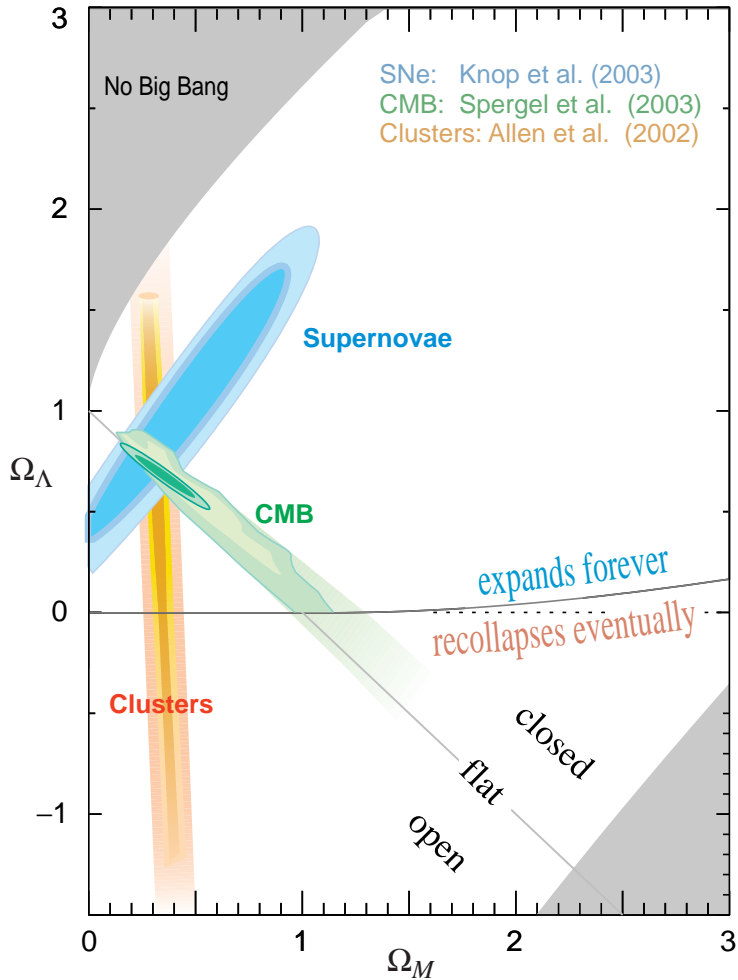}
\caption{
Summary plot showing the region
of sensitivities of the various
cosmology measurements in the
$ ( \Omega _m , \Omega _{\Lambda}  )$  plane.
Adapted from Ref.~\cite{pdgcosmos}.
}
\label{cosmos}
\end{figure}

Precision measurements of the cosmological
parameters is a subject where tremendous
amount of progress were made in the last
decade~\cite{pdgcosmos}, through the 
studies of cosmic microwave background,
large red-shift supernovae, as well as
the structures in galaxy clustering.
The summary plot in the 
$( \Omega _m , \Omega _{\Lambda} )$
plane is displayed in Figure~\ref{cosmos},
where $\Omega _m$ is the total mass density,
while $\Omega _{\Lambda}$ is the cosmological
constant. The best-fit allowed region is
consistent with a flat Universe 
( $ \Omega _m + \Omega _{\Lambda} \approx 1 $ )
with about $\Omega \sim 25\%$ 
in the form of known
or yet-unobserved matter, 
and  the remaining 
$\Omega \sim 75\%$ as ``Dark Energy''.

The total neutrino mass density affects the
scale of structure formation on the galaxy clusters.
Combined global analyses of the different probes
give rise
to upper bound on total neutrino mass
of $\sum m_i < 0.17 ~ \rm{eV}$ at 95\%
confidence level~\cite{cosmosnumass}.
Coupling with results from
neutrino oscillation experiments,
lower bound of neutrino mass density
of $\Omega _{\nu} > 0.001$ can be
placed.

\section{Neutrino Physics :  FUTURE }

The main thrust of neutrino 
physics will be to perform
precision studies on the already
measured parameters,
to uncover the
remaining parameters ($\theta _{13} , \delta$)
in {\it U}, to establish an mass scale
for $m_i$, to differentiate the two 
possible mass hierarchies, and to 
identify neutrinos as being Dirac or
Majorana particles.
In addition, among the various neutrino sources
depicted in Figure~\ref{nusource}, only
a relatively small window from $\sim$1~MeV to
$\sim$100~GeV is detectable by present techniques.
Research efforts towards
discovering and studying and devising
new neutrino sources, interaction channels 
as well as detection mechanisms remain 
crucial and complementary.
The future of neutrino experiments
will therefore evolve along 
the various directions:

\subsection{Oscillation Parameters:}

There is a running experiment with
accelerator neutrinos from the
Fermilab NuMI beam to the MINOS
detector 730~km away.
First results~\cite{minos06} are consistent
with interpretations of neutrino oscillation
at the atmospheric parameter space of
Eq.~\ref{eqn::atm}.
The CERN neutrino beam line is
under construction, and will be received by
the ICARUS and OPERA experiments at Gran Sasso,
also 730~km away.
Another approved project 
is the T2K experiment,
where an intense neutrino beam in Tokai, Japan
is being built to be sent to the Super-Kamiokande
detector 295~km away in an ``off-axis'' mode.
An alternative project NO$\nu$A is being pursued
using the Fermilab NuMI beam.
The goals of these experiments will be to improve on the 
($\Delta m^2_{23} ,  \theta _{23}$) 
measurements, to detect
$\nu_{\tau}$ appearance explicitly,
to confirm $\nu_{\mu} \rightarrow \nu_{\tau}$
oscillations, and to measure $\theta _{13}$
by searching for 
$\nu _{\mu} \rightarrow \nu _e$
oscillations, and to possibly probe
the neutrino mass hierarchies.

If $\theta _{13}$ is finite and 
the mixing is at least at the 1\% level,
the CP violating effects in the neutrino
sector can be studied, for instance, by
comparing the difference between
oscillation probabilities 
P($\nu_{e} \rightarrow \nu_{\mu}$)
and  
P($\nuebar \rightarrow \bar{\nu}_{\mu}$).
To pursue this program, one needs intense
neutrino beams and Mega-ton size detector.
There are intense efforts in devising
neutrino ``factory'', either with
improved conventional proton-on-target techniques,
or as decay products from muons or radioactive
$\beta$-decay isotopes produced in accelerators.

Complementary to accelerator-based efforts,
there are several reactor-based projects with
goals to search for $\theta _{13}$ in reactor
with about 1~km baseline, by looking for
$\nuebar$ disappearing through the comparison
of the neutrino spectra at this distance to
those at production point.

\subsection{Intrinsic Neutrino Properties:}

\begin{figure}
\centering
\includegraphics[width=10cm]{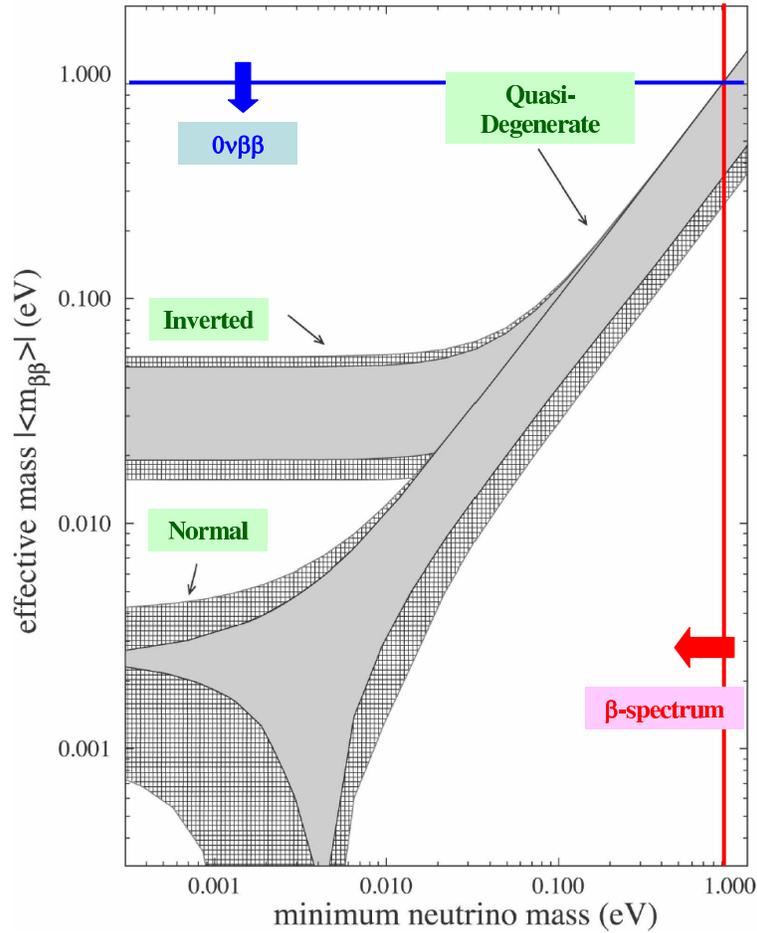}
\caption{
Sensitivity of the double beta decay
$\langle m_{\beta \beta} \rangle$ parameter
versus the minimum neutrino mass. The
allowed region from oscillation experiments
are shown by the bands. The inner
and outer bands
are due to the best-fit
oscillation parameters
and with 1$\sigma$ errors included,
respectively. 
Adapted from Ref.~\cite{vogelpdg}.
}
\label{0nubb}
\end{figure}


Distortions of the end-point of the $\beta$-decay
spectra provides direct measurements of 
the neutrino masses $m_i$'s 
and mixings $U_{ei} U_{ei}^*$.
The KATRIN experiment under construction will
push the sensitivity of to 0.2~eV,
and therefore can probe 
the scenario with
quasi-degenerate neutrino mass eigenstates,
as illustrated in Figure~\ref{0nubb}.
This sensitivity cannot resolve individual
$m_i$ and only the effective quantity ``electron
neutrino mass''
\begin{equation}
 m_{\nuebar} ^2  =
\sum_i | U_{ei} | ^2 m_i^2 
\end{equation}
can be determined or constrained.

Neutrinoless Double Beta Decay is 
the most promising avenue to
identify the Dirac or Majorana nature of
neutrinos, since this process is 
possible only for Majorana neutrinos.
The decay rates depend on an ``effective''
mass parameter :
\begin{equation}
\langle m_{\beta \beta} \rangle ^2  = 
\left| ~ | U_{e1} | ^2  m_1 
 +  e^{i \alpha } | U_{e2} |^2 m_2 
 +  e^{i \beta } | U_{e3} |^2  m_3 
~ \right| ^2
\end{equation}
and hence are sensitive
to the absolute scale of $\rm{m_i^2}$ rather
then $\rm{\Delta m^2 _{ij}}$. 
As depicted in Figure~\ref{0nubb},
when coupled with the input from oscillation
results, this processes can help
to distinguish the two possible mass
hierarchies.
Many R\&D projects on various candidate isotopes
are being pursued with the eventual goals
of achieving the interesting
range of 
$\langle m_{\beta \beta} \rangle \sim 0.1 - 0.01 ~ \rm{eV}$,
which covers the case of neutrino mass
spectrum having the inverted hierarchy,
as shown in Figure~\ref{0nubb}.

Finite neutrino masses typically give rise
to anomalous neutrino interactions not accounted
for by the Standard Model. The primary example is
the possible couplings between neutrinos
and photons via  their spin components, manifested
as neutrino magnetic moments and neutrino radiative
decays~\cite{mmreview}. 
Besides the model-dependent
astrophysical bounds, the
most stringent laboratory direct limits come from
reactor neutrino experiments~\cite{mmtexono}, at
$\mu _{\nu} ( \bar{\nu}_e ) < 7.4 \times 10^{-11} ~ \mu _{\rm B}$.
Bounds are also placed for the 
$\nu_e$-properties~\cite{rnue} 
as well as on axion emissions~\cite{raxion} from 
the reactor experiment.

\subsection{Detection of Weak/Rare Signals:}

The KamLAND experiment, with its
excellent sensitivity to $\nuebar$, 
has recently made the first observation of
the terrestrial ``geoneutrino''~\cite{geonu}
which are by-products of the $\beta$-decays
due to radioactivity in the Earth's crust,
pre-dominantly from the $^{238}$U and
$^{232}$Th series.
The various big underground experiments are
sensitive to neutrinos from supernovae, with
the expectation of 
detecting thousands of events from
the next supernova, as compared to the 20~events
from SN1987a. 

\subsection{High Energy Astrophysical and Cosmological Neutrinos:}

There are several ``neutrino telescope'' experiments~\cite{nu04}
(Lake Baikal, AMANDA, NESTOR, ANTARES, IceCube) 
based on the water or ice Cerenkov detection
techniques.
Their scientific goals are 
(a) to identify and
understand the high-energy 
($10^{12}$ to $10^{15}$~eV)
neutrino sources from active galactic nuclei, gamma-ray
bursts, neutron stars and other astrophysical
objects, and (b) to use these high-energy neutrinos
for neutrino physics like very long baseline studies.
The eventual detection volume for projects like IceCube
will be on the scale of 1~km$^3$.

To get above the $10^{18}$~eV scale for the ``GZK'' or
other neutrinos of cosmological origins, various
techniques are being pursued towards the detection
of radio Cerenkov, fluorescent, and acoustic signals
from the interaction showers~\cite{uhenu}.
A wide spectrum of detection media is be studied,
from ice to sea water to salt mines to the Moon 
to sampling a big region of the Earth's atmosphere
from Space.

\subsection{Neutrinos at Low Energy Frontiers:}

The Borexino and KamLAND experiments
will try to measure
the sub-MeV solar $^7$Be neutrinos,
while several R\&D projects are under way
to devise techniques
based on $\nu_e$N-charged current interactions
to detect the solar pp neutrinos.

Neutrino nucleus coherent scatterings have not
been experimentally observed. 
A sub-keV threshold detector of kg-size is
required to meet this challenge with reactor neutrinos.
To realize this goal,
prototype ``Ultra-Low-Energy'' germanium
detectors with software pulse shape discrimination
techniques are being developed~\cite{ulege},
where a threshold of O(100~eV) has been demonstrated.

Finally, the relic ``Big Bang'' neutrino, the
counterpart to the 2.7~K cosmic microwave photon background (CMB),
has large and comparable
number density (order of 100~$\rm{cm^{-3}}$)
but extremely small cross sections due to
the meV energy scale at an effective temperature of 1.9~K.
The relic neutrinos decouples from matter
at a much earlier time (1~s) than the CMB (3$\times 10^5$~years),
and hence are, in principle, better probes to the
early Universe.
A demonstration of its existence and a measurement
of its density is a subject of extraordinary
importance. Though there is no
realistic proposals on how to detect them,
it follows the traditions of 
offering a highly rewarding challenge to
and pushing the ingenuity of 
neutrino experimentalists.

\section{Outlook}

Neutrino physics and astrophysics will remain
a central subject in experimental particle physics
in the coming decades and beyond. 
The structures of the neutrino mass spectrum
and the mixing matrix will be studied
more thoroughly, as will be the intrinsic
neutrino properties as well as the high
and low energy frontiers.
There are much room for ground-breaking 
technical innovations -
as well as potentials for {\it further}
surprises.

\section*{Acknowledgments}

The author is grateful to the hospitality of the
ISSSMB06 organizers for the thoughtful arrangement
and the superb setting of the summer school, and to
the participants whose enthusiasm makes it 
so memorable. They are to be congratulated for
this success, and are encouraged to turn this into
a regular national, regional, or even international event.  
The work is supported by  
contract 94-2112-M-001-028 
from the National Science Council, Taiwan.


\begin{thebibliography}{99}
\bibitem{pdg}
``Review of Particle Physics'', Particle Data Group,
http://pdg.lbl.gov/ ,
J. Phys. {\bf G 33} (2006).
\bibitem{book}
For textbooks, see, for example:\\
``Neutrino Physics'', ed. K. Winter,
Cambridge University Press (1991);\\
``Physics of Massive Neutrinos'', 2nd Edition,
F. Boehm and P. Vogel, Cambridge University Press (1992);\\
``Neutrino Physics'', K. Zuber, Inst. of Phys. Press (2004).
\bibitem{nu04}
For latest status, see, for example:\\
``Proc. of the XXIst Int. Conf. on Neutrino Phys. \&
Astrophysics'', 
eds. J. Dumarchez, Th. Patzak and F. Vannucci,
Nucl. Phys. {\bf B} (Procs. Suppl.) {\bf 143} (2005).
\bibitem{weblink}
For~Web-portals, click, for example:\\
``Neutrino Oscillation~Industry'' $-$
http://neutrinooscillation.org/ ;\\
``Neutrino Unbound'' $-$
http://www.nu.to.infn.it/ .
\bibitem{pdgcosmos}
O. Lahav and A.R. Liddle, in Ref.~\cite{pdg}, 
J. Phys. {\bf G 33}, 224 (2006),
and references therein.
\bibitem{reinescowan}
F. Reines and C.L. Cowan, Phys. Rev. {\bf 90}, 492 (1953).
\bibitem{kayserpdg}
B. Kayser, in Ref.~\cite{pdg}, 
J. Phys. {\bf G 33}, 156 (2006),
and references therein.
\bibitem{vogelpdg}
P. Vogel and A. Piepke, in Ref.~\cite{pdg}, 
J. Phys. {\bf G 33}, 479 (2006),
and references therein.
\bibitem{msw}
L. Wolfenstein, Phys. Rev. {\bf D 17}, 2369 (1978);
S.P. Mikheyev and A. Yu. Smirnov, Sov. J. Nucl. Phys.
{\bf 42}, 1441 (1985).
\bibitem{superk}
Y. Ashie et al., Super-Kamiokande Coll.,
Phys. Rev. {\bf D 71}, 112005 (2005), and
references therein. 
\bibitem{k2k}
E. Aliu et al, K2K Coll. 
Phys. Rev. Lett. {\bf 94}, 081802 (2005).
\bibitem{bahcallbook}
``Neutrino Astrophysics'', J.N. Bahcall, Cambridge University
Press (1989).
\bibitem{sno}
Q.R. Ahmad et al. SNO Coll., Phys. Rev. Lett. {\bf 89} 011301 (2002);
B. Aharmim et al. SNO Coll., Phys. Rev. {\bf C 72}, 
055502 (2005). 
\bibitem{solarmodel}
J.N. Bahcall, A.M. Serenelli and S. Basu, Astrophys. J. 
{\bf 621}, L85 (2005),
and references therein. 
\bibitem{kamland}
K. Eguchi et al., KamLAND Coll.,
Phys. Rev. Lett. {\bf 90}, 021802 (2003).
T. Araki et al., KamLAND Coll.,
Phys. Rev. Lett. {\bf 94}, 081801 (2005).
\bibitem{lsnd}
A. Aguilar et al., LSND Coll.,
Phys. Rev. {\bf D 64}, 112007 (2001).
\bibitem{chooz}
M. Apollonio et al., CHOOZ Coll., Phys. Lett. {\bf B 466}, 415 (1999).
\bibitem{pdgbbn}
B.D. Fields and S. Sarkar, in Ref.~\cite{pdg}, 
J. Phys. {\bf G 33}, 220 (2006),
and references therein.
\bibitem{nufamily}
R.H.~Cyburt et al., Astropart. Phys. {\bf 23}, 313 (2005).
\bibitem{cosmosnumass}
U. Seljak, A. Slosar and P. McDonald,
JCAP {\bf 0610}, 014 (2006).
\bibitem{minos06}
D.G.~Michael et al., MINOS Coll., 
Phys. Rev. Lett. {\bf 97}, 191801 (2006).
\bibitem{mmreview}
H.T.~Wong, Mod. Phys. Lett. A 20, 1103 (2005),
and references therein.
\bibitem{mmtexono}
H.T.~Wong et al., TEXONO Coll.,
Phys. Rev. {\bf D 75}, 012001 (2007).
\bibitem{rnue}
B.~Xin et al., TEXONO Coll.,
Phys. Rev. {\bf D 72}, 012006 (2005).
\bibitem{raxion}
H.M.~Chang et al., TEXONO Coll.,
hep-ex/0609001,
Phys. Rev. {\bf D}, in press (2007).
\bibitem{geonu}
T. Araki et al., KamLAND Coll., Nature {\bf 436},
499 (2005).
\bibitem{uhenu}
R. Nahnhauer, in Ref.~\cite{nu04},
Nucl. Phys. {\bf B} (Procs. Suppl.) {\bf 143}, 387 (2005),
and references therein.
\bibitem{ulege}
H.T.~Wong et al., J. Phys. Conf. Ser. {\bf 39}, 266 (2006)
\end{thebibliography}
\end{document}